%% file: HIG-19-013_temp.tex
\pdfoutput=1

\documentclass[11pt,twoside,a4paper,cmspaper,final,collab]{cms-tdr}
\def\svnVersion{1c16987-D}\def\svnDate{2020/08/10}\def\cmsCernNoTag{CERN-EP-2020-028}\def\cmsCernDate{\today}\def\cmsMessage{"Published in Physical Review Letters as \href{http://dx.doi.org/10.1103/PhysRevLett.125.061801}{\doi{10.1103/PhysRevLett.125.061801}.}"}

\begin{document}\cmsNoteHeader{HIG-19-013}

\hyphenation{had-ron-i-za-tion}
\hyphenation{cal-or-i-me-ter}
\hyphenation{de-vices}
\newcommand{\aMCATNLO} {\textsc{MG5}\_a\MCATNLO\xspace}
\newcommand{\jhugen}{\textsc{JHUGen}\xspace}
\newcommand{\ttH}{\ensuremath{\ttbar\PH}\xspace}
\newcommand{\Htt}{\ensuremath{\PH\PQt\PQt}\xspace}
\newcommand{\Hff}{\ensuremath{\PH\text{ff}}\xspace}
\newcommand{\VH}{\ensuremath{\PV\PH}\xspace}
\newcommand{\VBF}{\ensuremath{\mathrm{VBF}}\xspace}
\newcommand{\bbH}{\ensuremath{\bbbar\PH}\xspace}
\newcommand{\tH}{\ensuremath{\cPqt\PH}\xspace}
\newcommand{\ggH}{\ensuremath{\cPg\cPg\PH}\xspace}
\newcommand{\VV}{\ensuremath{\PV\PV}\xspace}
\newcommand{\HVV}{\ensuremath{\PH\PV\PV}\xspace}
\newcommand{\mH}{\ensuremath{m_{\PH}}\xspace}
\newcommand{\fcp}{\ensuremath{f_{\mathrm{CP}}^{\Htt}}\xspace}
\newcommand{\Hgammagamma}{\ensuremath{\PH \to \PGg\PGg}\xspace}
\newcommand{\gammagamma}{\ensuremath{\PGg\PGg}\xspace}
\newcommand{\mgg}{\ensuremath{m_{\PGg\PGg}}\xspace}
\newcommand{\DYjets}{\ensuremath{\PZ/\PGg^{*}\text{+jets}}\xspace}
\newcommand{\ttbarr}{\ensuremath{\ttbar + \text{jets}}\xspace}
\newcommand{\ttgammagamma}{\ensuremath{\ttbar + \PGg\PGg}\xspace}
\newcommand{\ttgamma}{\ensuremath{\ttbar + \PGg}\xspace}
\newcommand{\qcd}{\ensuremath{\text{jet} + \text{jet}}\xspace}
\newcommand{\gjet}{\ensuremath{\PGg + \text{jets}}\xspace}
\newcommand{\dipho}{\ensuremath{\PGg\PGg + \text{jets}}\xspace}
\newcommand{\vgamma}{\ensuremath{\PV + \PGg}\xspace}
\newcommand{\zgamma}{\ensuremath{\PZ + \PGg}\xspace}
\newcommand{\wgamma}{\ensuremath{\PW + \PGg}\xspace}
\newcommand{\sigtimesbr}{\ensuremath{\sigma_{\ttH} \mathcal{B}_{\PGg\PGg}}}
\ifthenelse{\boolean{cms@external}}{\providecommand{\tablewidth}{1\linewidth}}{\providecommand{\tablewidth}{0.8\linewidth}}
\ifthenelse{\boolean{cms@external}}{\providecommand{\figuretwowidth}{0.48\textwidth}}{\providecommand{\figuretwowidth}{0.6\textwidth}}
\ifthenelse{\boolean{cms@external}}{\providecommand{\figurethreewidth}{0.48\textwidth}}{\providecommand{\figurethreewidth}{0.55\textwidth}}
\ifthenelse{\boolean{cms@external}}{\providecommand{\cmsTable}[1]{#1}}{\providecommand{\cmsTable}[1]{\resizebox{\textwidth}{!}{#1}}}
\ifthenelse{\boolean{cms@external}}{\providecommand{\CLnp}{C.L}}{\providecommand{\CLnp}{CL}}
\ifthenelse{\boolean{cms@external}}{\renewcommand{\CL}{C.L.}\xspace}{\renewcommand{\CL}{CL\xspace}}

\cmsNoteHeader{HIG-19-013}
\title{Measurements of \texorpdfstring{$\ttH$}{ttbar H} production and the CP structure of the Yukawa interaction between the Higgs boson and the top quark in the diphoton decay channel}

\date{\today}

\abstract{
The first observation of the \ttH process in a single Higgs boson decay channel with the full reconstruction of the final state ($\Hgammagamma$) is presented, with a significance of 6.6 standard deviations ($\sigma$). The CP structure
of Higgs boson couplings to fermions is measured, resulting in an exclusion of the pure CP-odd structure of the top Yukawa coupling
at $3.2\sigma$. The measurements are based on a sample of proton-proton collisions at a center-of-mass energy $\sqrt{s}=13$\TeV
collected by the CMS detector at the LHC, corresponding to an integrated luminosity of 137\fbinv.
The cross section times branching fraction of the \ttH process is measured to be $\sigtimesbr= 
1.56^{+0.34}_{-0.32}$\unit{fb}, which is compatible with the standard model prediction of $1.13^{+0.08}_{-0.11}$\unit{fb}.
The fractional contribution of the CP-odd component is measured to be $\fcp=0.00\pm0.33$.
}

\hypersetup{
pdfauthor={CMS Collaboration},
pdftitle={Measurements of ttbar H production and the CP structure of the Yukawa interaction between the Higgs boson and the top quark in the diphoton decay channel},
pdfsubject={CMS},
pdfkeywords={CMS, physics, Higgs, top quarks}}

\maketitle

Since its observation~\cite{Aad:2012tfa, Chatrchyan:2012xdj, Chatrchyan:2013lba},
the properties of the Higgs boson (\PH) have been studied using a variety of decay channels and production modes.
Among these properties, the tree-level top quark Yukawa (\Htt) coupling and its CP structure can be tested by 
studying \PH production in association with a top quark-antiquark pair (\ttH).
The CMS~\cite{Sirunyan:2018hoz} and ATLAS~\cite{Aaboud:2018urx} Collaborations reported the observation
of the \ttH process by combining several \PH decay channels, with a cross section
compatible with the standard model (SM) expectation.
One of the most sensitive channels for probing the \ttH process is \Hgammagamma.
By probing the interaction between the \PH and vector bosons,
CMS~\cite{Chatrchyan:2012jja,Chatrchyan:2013mxa,Khachatryan:2014kca,Khachatryan:2015mma,Khachatryan:2016tnr,Sirunyan:2017tqd,Sirunyan:2019twz,Sirunyan:2019nbs}
and ATLAS~\cite{Aad:2013xqa,Aad:2015mxa,Aad:2016nal,Aaboud:2017oem,Aaboud:2017vzb,Aaboud:2018xdt}
have determined that the \PH quantum numbers are consistent with $J^{PC}=0^{++}$.
However, small anomalous contributions were not excluded, and
studies of the \Htt coupling provide an alternative and independent path for CP tests in the Higgs sector~\cite{Gunion:1996xu,Demartin:2014fia,Gritsan:2016hjl}. 

This Letter reports on the measurement of the production rate of \ttH
with \Hgammagamma, giving the first observation of the tree-level \Htt coupling in a single \PH decay channel,
along with a first test of its CP structure. Results are based on data from
proton-proton ($\Pp\Pp$) collisions at a center-of-mass energy of
$\sqrt{s}=13$\TeV collected with the CMS detector at the LHC between 2016 and 2018, corresponding to an integrated luminosity of 137\fbinv. 

The central feature of the CMS apparatus \cite{Chatrchyan:2008zzk}
is a superconducting solenoid of 6\unit{m} internal diameter,
providing a magnetic field of 3.8\unit{T}. Inside the solenoid there is a silicon
tracker, a lead-tungstate crystal electromagnetic calorimeter (ECAL), and a brass and scintillator hadron calorimeter.
Forward calorimeters extend the coverage to higher pseudorapidity ($\eta$),
and muon detectors are embedded in the flux-return yoke of the solenoid.

The particle-flow (PF) algorithm~\cite{CMS-PRF-14-001}
reconstructs individual particles (photons,
charged and neutral hadrons, muons, and electrons)
by combining information from all detectors.
Jets are built from PF particles with the anti-\kt algorithm~\cite{Cacciari:2008gp,Cacciari:2011ma}
with a distance parameter of 0.4.
The missing transverse momentum ($\ptmiss$) is defined as the negative vector sum of the transverse momenta ($\pt$) of all PF
particles. The primary $\Pp\Pp$ interaction vertex is taken as the vertex with the largest value of summed physics-object $\pt^{2}$~\cite{CMS-TDR-15-02}.
Charged hadrons originating from additional $\Pp\Pp$  interactions are removed
from the analysis. Jets from the hadronization of bottom
quarks are tagged by a secondary vertex algorithm 
based on the score from a deep neural network (DNN) \cite{Sirunyan:2017ezt}.

Signal and background processes are generated with several Monte Carlo (MC) programs.
All \PH production processes are modeled with \MGvATNLO 2.4.2 at next-to-leading order (NLO) \cite{Alwall:2014hca}
in quantum chromodynamics (QCD), with cross sections and decay branching fractions
taken from Ref.~\cite{deFlorian:2016spz}. 
A separate \ttH sample, generated with \textsc{Powheg 2.0}~\cite{Nason:2004rx,Frixione:2007vw,Alioli:2010xd,Hartanto:2015uka} 
at NLO in QCD, is used to increase the number of events used for training the multivariate discriminants described below.
For the CP study, \ttH anomalous coupling samples of CP-odd, CP-even, and a mixture of the two, are generated
at leading order (LO) with \jhugen~7.0.2~\cite{Gao:2010qx,Bolognesi:2012mm,Anderson:2013afp,Gritsan:2016hjl}
and reweighted with the \textsc{MELA} matrix element library~\cite{Gao:2010qx,Bolognesi:2012mm,Anderson:2013afp,Gritsan:2016hjl}, 
\jhugen~7.0.2 and \textsc{MELA} are also used for the study of CP effects in the $\cPqt\PH$ process. 
The \MGvATNLO program is also used to generate most background processes, e.g.,
\ttgammagamma, \ttgamma, \ttbarr, \gjet, \vgamma, Drell--Yan, diboson, $\cPqt + \PV$, where \PV is a \PW or a \PZ boson.
In contrast, the diphoton background (\dipho) is generated with \textsc{sherpa}
2.2.4~\cite{10.21468/SciPostPhys.7.3.034}, which
includes tree-level processes with up to three additional jets, as well as box processes at LO accuracy.
In all MC samples, the parton fragmentation and hadronization and the underlying events
are modeled with \PYTHIA 8.205~\cite{Sjostrand:2014zea}, with the CUETP8M1~\cite{Khachatryan:2015pea} and CP5~\cite{Sirunyan:2019dfx}
tune used for the simulation of 2016 and 2017/2018 data respectively.
Finally, the detector response is simulated with the \GEANTfour package \cite{Agostinelli:2002hh}.

The trigger~\cite{Khachatryan:2016bia} selects diphoton events with a loose calorimetric identification~\cite{Sirunyan:2018ouh},
and asymmetric photon transverse energy ($\ET$) thresholds
of 30 and 18 (22)\GeV for the data collected during 2016 (2017/2018).
The trigger efficiency is $>$95\% and is measured 
as a function of $\ET$, $\eta$, and $\RNINE$ of the photons using an alternative trigger,
where $\RNINE$ is the energy sum of the $3\times3$ crystals centered around the most energetic crystal in the
cluster divided by the energy of the photon.

\PH candidates are built from pairs of photon candidates, which
are reconstructed from energy clusters in the ECAL not linked to 
charged-particle tracks (with the exception of converted photons).
The photon energies are corrected for the containment of electromagnetic showers
in the clustered crystals and the energy losses of converted photons
with a multivariate regression technique based on simulation~\cite{Sirunyan:2018ouh}.
The ECAL energy scale in data is corrected using 
$\PZ\to\EE$ simulated events smeared to reproduce the energy resolution measured in data.
The offline diphoton selection criteria are
similar to, but more stringent than, those used in the trigger~\cite{Sirunyan:2018ouh}.

Photons are further required to satisfy a loose identification (photon ID~\cite{Sirunyan:2018ouh})
criterion based on a boosted decision tree (BDT) classifier trained to separate photons from jets.
Inputs to photon ID such as shower shape and isolation variables in simulation are corrected with
a chained quantile regression method~\cite{DBLP:journals/corr/abs-1211-6581} based on studies of $\PZ\to\EE$ events.
Each variable is corrected with a separately trained BDT,
taking the photon kinematic properties, per-event energy density, and the previously corrected features as inputs, to ensure that correlations between the inputs are preserved and closer to those in data.  
This method improves the modeling of the photon ID BDT discriminant in MC simulation
with respect to the previous CMS \Hgammagamma results~\cite{Sirunyan:2018ouh}.

After the preselection described above, we require $100 < \mgg < 180$\GeV,
$\pt/\mgg>$ 1/3 and 1/4 for the leading (in \pt) and subleading photons respectively
and then divide events into two channels.
The leptonic channel is aimed at selecting events where at least one top quark decays
leptonically, and demands the presence of ${\geq}1$ jet with $\pt > 25$\GeV and $\abs{\eta} < 2.4$,
${\geq}1$ isolated \Pe or \Pgm with $\pt > 10$\GeV for electrons, $\pt > 5$\GeV for muons, and $\abs{\eta} < 2.4$. 
The hadronic channel targets \ttbar hadronic decays by
requiring at least three jets, 
at least one \cPqb-tagged jet, and no isolated leptons (\Pe/\Pgm). 

A dedicated BDT discriminant (``BDT-bkg'') is employed in each channel to distinguish between \ttH and
background events. These BDTs are trained with the \textsc{xgboost} \cite{xgboost} framework 
on signal and background MC samples, with one exception as noted below.
The background MC samples include \gjet, \dipho, \ttbarr, \ttgamma, \ttgammagamma, \zgamma, and \wgamma processes, 
as well as a variety of other rarer backgrounds. Non-\ttH production modes of \PH are also treated as background.
The dominant background in the hadronic channel consists of \gjet events, where one jet is misidentified as a photon.
To improve the performance of the hadronic BDT-bkg,
the \gjet background is modeled from a large sample of
data events with one photon candidate failing the photon ID requirement;
these are almost exclusively multi-jet and \gjet events.
For each such event, the photon ID value of the misidentified jet is replaced by a value drawn from the MC distribution
of photon ID values of misidentified jets passing the photon ID requirement.
These events, appropriately weighted, are then used in the hadronic BDT-bkg training instead
of the \gjet MC sample.

\begin{figure*} [htb!]
    \centering
    \includegraphics[width=0.48\textwidth]{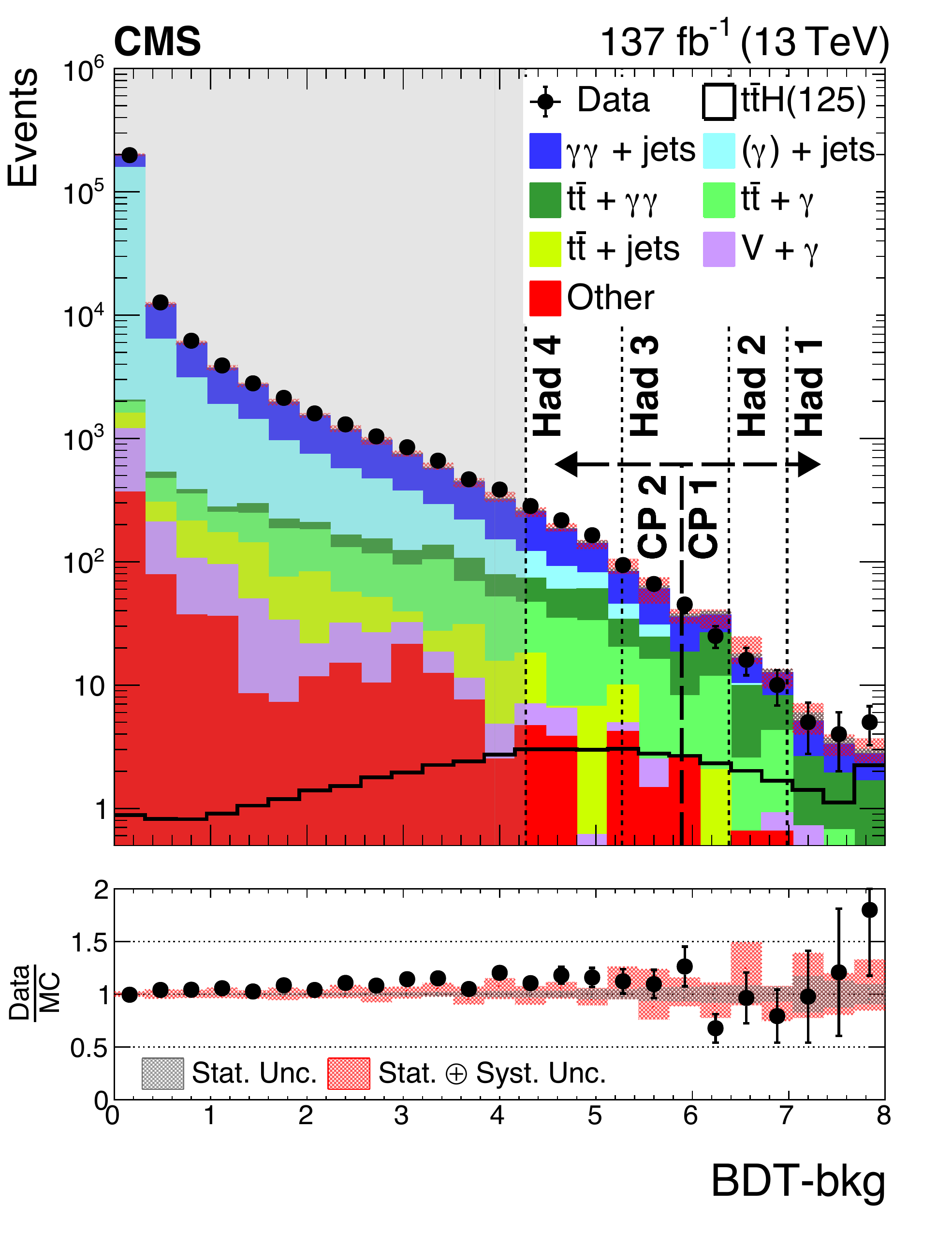} 
    \includegraphics[width=0.48\textwidth]{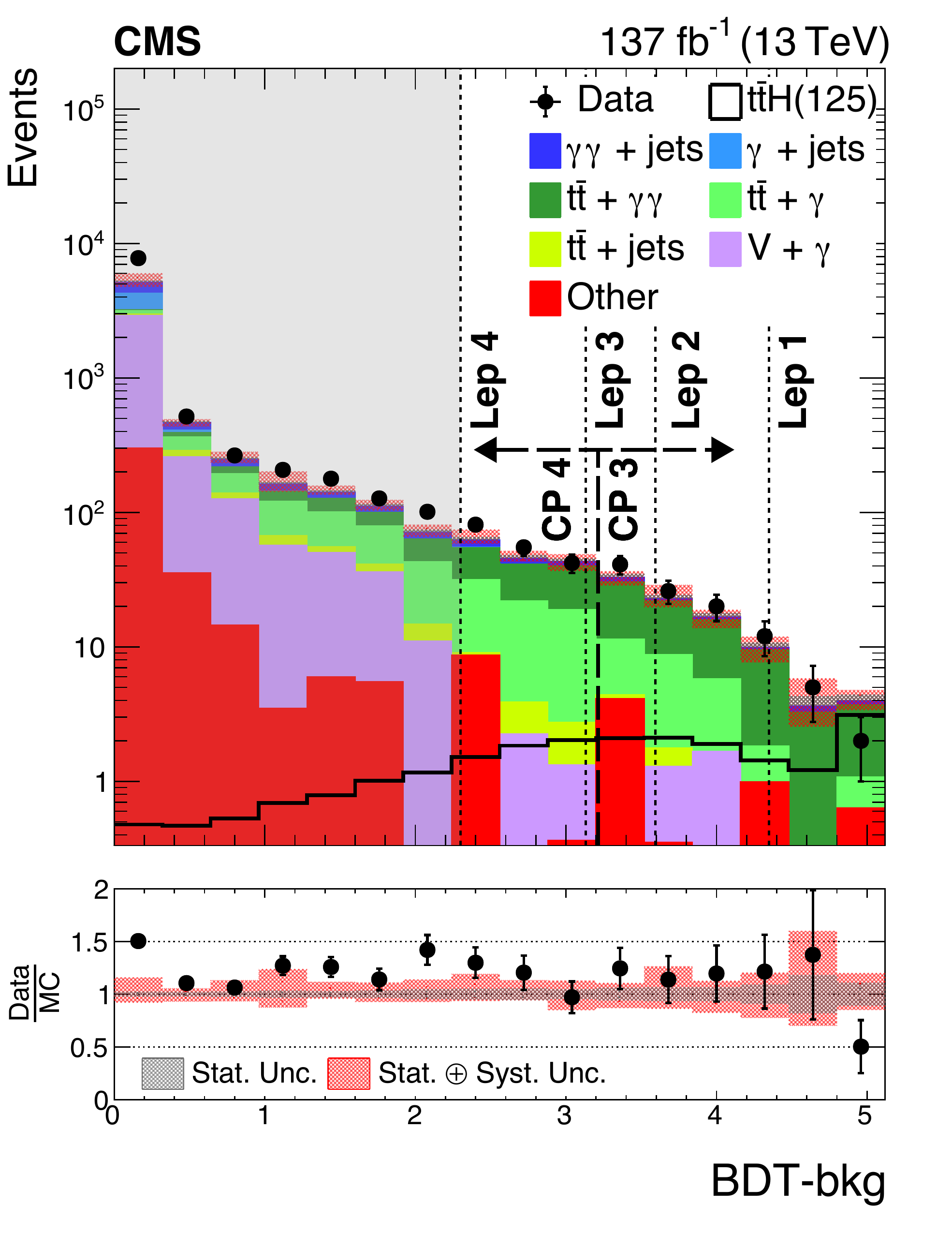}
    \caption{Distributions of BDT-bkg output used for event categorization,
             for the hadronic (left) and the leptonic (right) channels.
             Category boundaries for the signal strength (CP) measurements are shown
             with thinly (thickly) dashed lines. Events shown are taken from the
             \mgg sidebands, satisfying either $100 < \mgg < 120$\GeV
             or $130 < \mgg < 180$\GeV. Events in the grey shaded region are not
             considered in the analysis. Statistical (statistical $\oplus$ systematic) background 
             uncertainties are represented by the black (red) shaded bands.
    }
    \label{fig:bdt_score}
\end{figure*}

Input features of BDT-bkg include kinematic properties of jets,
leptons, photons and diphotons (but not \mgg),
jet and lepton multiplicity, \cPqb-tagging scores of jets, and \ptmiss.
The inclusion of \cPqb-tagging scores reduces the non-\ttbar background; furthermore,
jets and leptons in \ttH events tend to have higher $\pt$
and smaller $\abs{\eta}$ than in background events.
The BDT-bkg also uses output of the photon ID BDT,
and the outputs of other machine learning (ML) algorithms described below as input features.
One such ML algorithm is a top quark tagger BDT (top tagger)~\cite{Sirunyan:2017wif} to distinguish events
with top quarks decaying into three jets from events that do not contain top quarks. 
We also use long short-term memory based \cite{HochSchm97} DNNs trained to separate \ttH
from the dominant backgrounds in a signal-enriched phase space:
\dipho and \ttgammagamma for the hadronic channel; and \ttgammagamma for the leptonic channel. 
In addition to the features that are used in BDT-bkg, the DNNs exploit low-level information
including the full four-vectors of each jet and lepton and the
jet flavor scores \cite{Sirunyan:2017ezt}. The four-vectors allow for a more effective use of
the kinematic properties of the jet and the lepton while the jet flavor scores allow the differentiation of the origins of hadronic
jets between \ttH and the dominant backgrounds that the DNNs are designed to reject.
The DNNs are trained only on MC samples with a large number of simulated events and 
used as additional inputs to the BDT-bkg, rather than in place of the BDT-bkg.
When a DNN is trained on all background components its performance is worse than the BDT-bkg due to severe overfitting,
as the other background samples have a lower number of simulated events than \dipho and \ttgammagamma.
The modeling of the input features has been validated by
comparing data and MC distributions for events passing the preselection in both channels.
The BDT-bkg score has been validated by comparing the distributions in data and MC in both
the \mgg sidebands, satisfying either $100 < \mgg < 120$\GeV or $130 < \mgg < 180$\GeV
(as in Fig.~\ref{fig:bdt_score}) as well as in dedicated control regions which target
$\ttbar + \PZ$ events.

Events are either rejected or further divided into eight categories to maximize the expected
significance according to their BDT-bkg output as shown in Fig. \ref{fig:bdt_score} and
Table~\ref{tab:results_SigBkgYields}. 
When measuring the CP structure of the \Htt coupling that is discussed later,
non-rejected events are divided into four categories 
to maximize the sensitivity to the CP structure of the \Htt amplitude.
We perform a simultaneous binned
maximum likelihood fit to the \mgg distributions
in the eight categories to extract the product of the \ttH cross section and $\Hgammagamma$ 
branching fraction (\sigtimesbr) and the signal strength
$\mu_{\ttH}$, defined as the ratio of the measured to SM expected \Hgammagamma.
In the fit, all other \PH production modes are constrained to their SM predictions.

The \ttH signal distribution is parameterized using a double-sided Crystal Ball~\cite{CrystalBallRef} plus Gaussian function.
The background is modeled from data with the discrete profiling method~\cite{Dauncey:2014xga},
which accounts for the uncertainty associated with the choice 
of analytic function used to model the background \mgg distribution.

All other systematic uncertainties are also included as nuisance parameters,
and results are obtained using asymptotic distributions of test statistics based on the profile likelihood ratio~\cite{CMS-NOTE-2011-005,Cowan:2010js,Khachatryan:2014jba}.
The dominant theoretical uncertainty in
$\mu_{\ttH}$ arises from the SM prediction of the \ttH cross section,
and is estimated by varying the QCD renormalization and factorization scales~\cite{deFlorian:2016spz}, with a resulting impact of 8\%.
The uncertainties in parton distribution functions, QCD coupling,
underlying event and parton showers, and the \Hgammagamma 
branching fraction each affect $\mu_{\ttH}$ by 2--5\%.
The main experimental uncertainties that affect $\mu_{\ttH}$ are those related to
the \cPqb quark and photon identification, the jet energy scale and resolution,
and the integrated luminosity~\cite{CMS-PAS-LUM-17-001,CMS-PAS-LUM-17-004,CMS-PAS-LUM-18-002}.  Their effects are in the 2--6\% range. 
Other systematic uncertainties, including those related to preselection and trigger efficiencies, the lepton identification, and \ptmiss
have a $<$2\% effect on the measurement of $\mu_{\ttH}$ and \sigtimesbr.

\begin{figure}[htbp]
    \centering
    \includegraphics[width=\figuretwowidth]{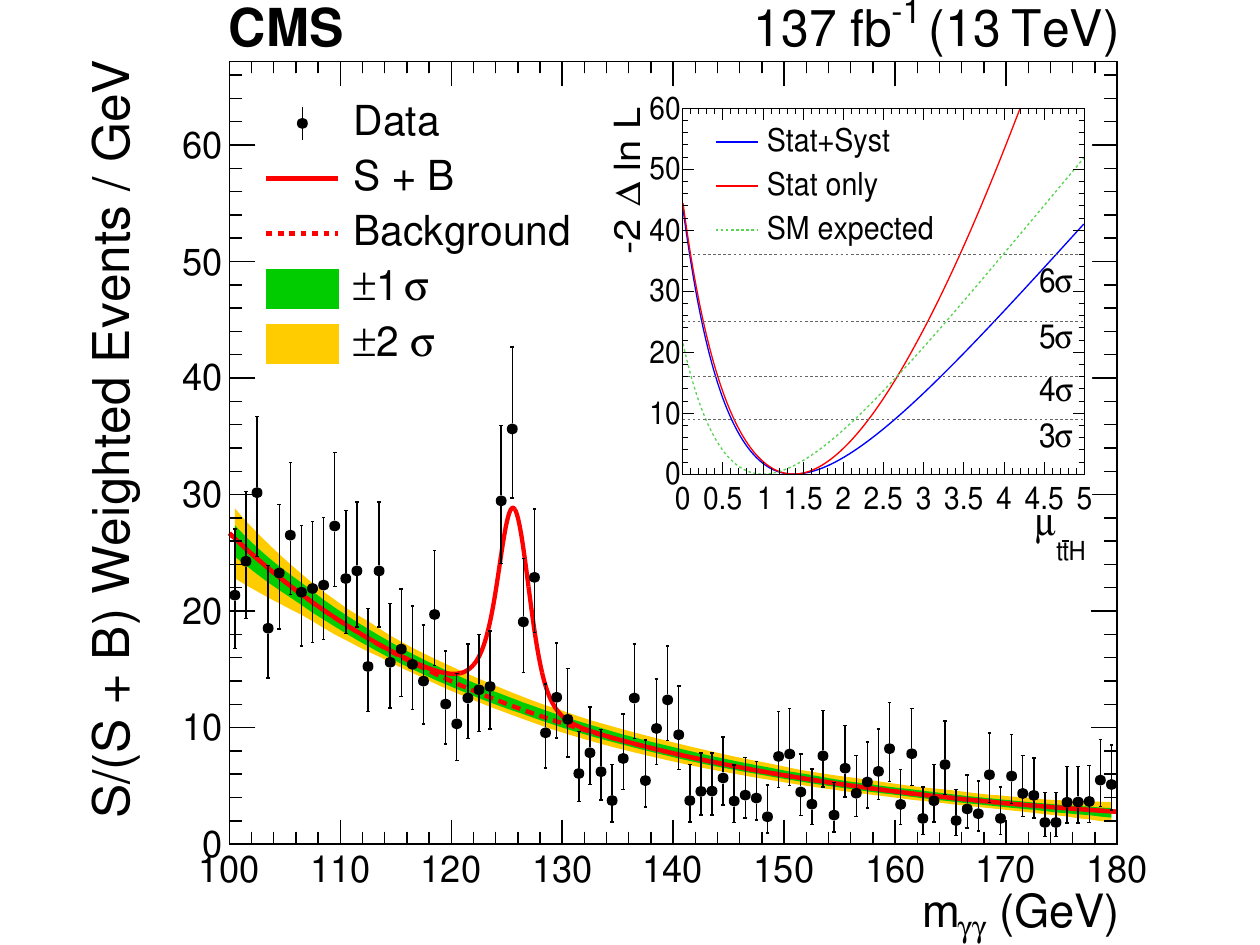}
    \caption{
        Invariant mass distribution for the selected events (black points) weighted by S/(S + B),
        where S (B) is the numbers of expected signal (background) events 
        in a $\pm 1\sigma_{\mathrm{eff}}$ mass window centered on \mH. 
        The $\sigma_{\mathrm{eff}}$ is defined as the smallest
        interval containing 68.3\% of the \mgg distribution,
        and ranges from 1.2 to 1.6\% for different categories.
        We show curves for fitted signal $+$ background (solid red) and for
        background only (dashed red), with bands covering the $\pm 1\sigma$ and $\pm 2\sigma$
        uncertainties in the fitted background. The inner panel shows the likelihood scan for $\mu_{\ttH}$ with \mH profiled.}
    \label{fig:results_observed_mu}
\end{figure}

The data and fit results are shown in Fig.~\ref{fig:results_observed_mu}.
We find $\sigtimesbr = 1.56^{+0.34}_{-0.32}$\unit{fb}$ = 1.56^{+0.33}_{-0.30} \stat$
$ ^{+0.09}_{-0.08} \syst$\unit{fb}, and
$\mu_{\ttH} = 1.38^{+0.36}_{-0.29} = 1.38^{+0.29}_{-0.27} \stat$
$ ^{+0.21}_{-0.11} \syst$ with the \PH mass (\mH) profiled. 
The SM prediction of the \sigtimesbr is  $1.13^{+0.08}_{-0.11}$\unit{fb} \cite{deFlorian:2016spz}.
The observed significance relative to the background-only hypothesis is
6.6 standard deviations ($\sigma$), while the expected significance assuming the SM \PH is $4.7\sigma$.

\begin{table*}
    \centering
    \topcaption{The expected number of \PH events in the hadronic and leptonic channels per category and the fractional contribution per \PH production mode.}
        \begin{scotch}{ r  c  c  c  c  c   c   c   c   c   c   c   c }
            &  Total & \ttH (\%) & \tH (\%) & \ggH (\%) & \VH (\%) & \VBF (\%) & \bbH (\%) \\  \hline
    Had1  &5.8  &89.1	&6.8	&3.3	&0.8	&$<$0.1	& 0.1\\	
    Had2  &4.2  &82.9	&6.8	&8.7	&1.4	& 0.2	& 0.1\\	
    Had3  &11.6 &78.6	&7.2	&10.3	&3.5	& 0.3	& 0.1\\	
    Had4  &13.6 &65.4	&7.7	&19.3	&6.9	& 0.7	& 0.1\\	
    Lep1  &5.8  &90.6	&7.9	&0.5	&1.0	&$<$0.1	&$<$0.1\\	
    Lep2  &4.9  &90.0	&6.7	&0.4	&2.9	&$<$0.1	&$<$0.1\\	
    Lep3  &3.5  &86.2	&7.4	&0.4	&6.0	&$<$0.1	&$<$0.1\\	
    Lep4  &5.7  &78.1	&8.2	&1.1	&12.7	&$<$0.1	&$<$0.1\\	
    Total &55.1 &79.5	&7.4	&8.2	&4.7	& 0.3	&$<$0.1\\	
    \end{scotch}
    \label{tab:results_SigBkgYields}
\end{table*}

The CP structure of the \Htt amplitude can be parameterized as~\cite{Gritsan:2016hjl}:
\begin{linenomath}
\begin{equation}
  \begin{aligned}
    && \mathcal{A}(\PH \PQt\PQt) = - \frac{m_{\PQt}}{v}
    \overline{\psi}_{\PQt} \left ( \kappa_{\PQt}  + \mathrm{i}  \tilde\kappa_{\PQt}  \gamma_5 \right ) {\psi}_{\PQt} ,
    \label{eq:ampl-spin0-qq} 
  \end{aligned}
\end{equation}
\end{linenomath}
where $\overline{\psi}_{\PQt}$ and ${\psi}_{\PQt}$ are the Dirac spinors, $m_{\PQt}$ is the top quark mass,
$v$ is the SM \PH field vacuum expectation value, and $\kappa_{\PQt}$ and $\tilde\kappa_{\PQt}$ are
the CP-even and CP-odd Yukawa couplings. In the SM, $\kappa_{\PQt}=1$ and $\tilde\kappa_{\PQt}=0$.
We measure the CP structure with
\begin{linenomath}
\begin{equation}
    \begin{aligned}
        \fcp = \frac{\abs{\tilde\kappa_\PQt}^2 }{\abs{\kappa_\PQt}^2 + \abs{\tilde\kappa_\PQt}^2}\ {\mathrm{sign}}(\tilde\kappa_\PQt / \kappa_\PQt).
\label{eq:fCP_definitions}
\end{aligned}
\end{equation}
\end{linenomath}
When the cross sections of the CP-even and CP-odd contributions are equal, $\fcp=0.72$~\cite{Gritsan:2016hjl}.

It has been shown in Ref.~\cite{Gritsan:2016hjl} that an optimal analysis of the CP structure in the \ttH process
can be performed with two observables, $\mathcal{D}_{0-}$ and $\mathcal{D}_{\mathrm{CP}}$.
$\mathcal{D}_{0-}$ is designed to separate CP-even from CP-odd and $\mathcal{D}_{\mathrm{CP}}$ to differentiate the interference.
Ref.~\cite{Gritsan:2020pib} shows that the two observables built by matrix element and ML techniques achieve the same sensitivity.
In this study, we use a BDT to obtain $\mathcal{D}_{0-}$ and do not include $\mathcal{D}_{\mathrm{CP}}$ since it requires tagging 
the flavor of light jets. As a consequence, it is not possible to measure the relative sign, or phase,
of the $\kappa_\PQt$ and $\tilde\kappa_\PQt$ couplings. Nonetheless, this sign is incorporated into the \fcp definition
in Eq.~\ref{eq:fCP_definitions} for consistency with other possible studies sensitive to the sign of \fcp,
such as in the gluon fusion production with the top-quark loop~\cite{Gritsan:2020pib}.

\begin{figure}[htbp]
    \centering
\includegraphics[width=\figurethreewidth]{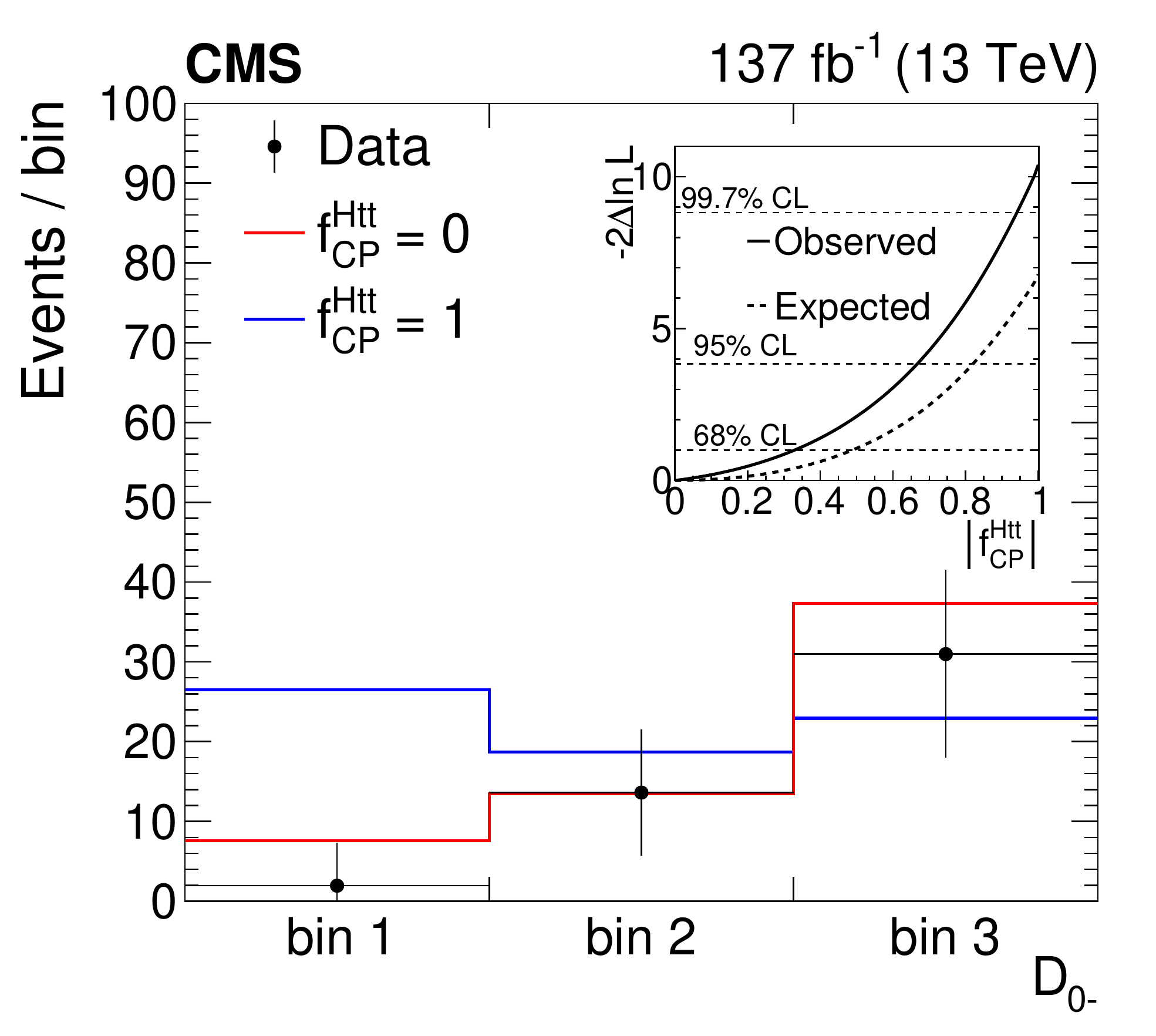}
\caption{
The distribution of events weighted by S/(S + B), as in Fig.~\ref{fig:results_observed_mu},
in three bins of the  $\mathcal{D}_{0-}$ discriminant. In this display, leptonic/hadronic channels
and BDT-bkg categories are combined in the mass range $115<\mgg<135$\GeV and
the background contribution, as determined in the fit to data, is subtracted.
The inner panel shows the likelihood scan for $\abs{\fcp}$.
}
\label{fig:bdt_cp}
\end{figure}

We train a BDT to distinguish CP-even and CP-odd contributions. 
The observables used in the training include the kinematic variables of the first six jets (in \pt)
and the diphoton system (but not \mgg), the \cPqb-tagging scores of jets, and in the leptonic channel, 
the lepton multiplicity and the kinematic variables of the leading lepton.
The output of the BDT is the $\mathcal{D}_{0-}$ observable. Simulation shows that $\mathcal{D}_{0-}$  
has negligible correlation with the BDT-bkg discriminant. 
The events selected for the signal strength measurements are split into 12 categories,
leptonic or hadronic, two BDT-bkg categories shown in Fig.~\ref{fig:bdt_score}, and three $\mathcal{D}_{0-}$ bins, 
as shown in Fig.~\ref{fig:bdt_cp}.

A simultaneous fit to the \mgg distribution is performed using the 12 categories 
to measure \fcp. The $\mu_{\ttH}$ parameter is left unconstrained. 
An additional systematic uncertainty is introduced to cover possible small differences in the modeling 
of the distributions with the \jhugen generator used for variation of the CP structure of the
\ttH coupling and \MGvATNLO generator used to model SM distributions. However,
statistical uncertainties dominate the measurement of \fcp. 
In addition to the \ttH process, we parameterize the $\cPqt\PH$ production with the $\mu_{\ttH}$ and
\fcp parameters, where the \PH couplings to other particles are constrained to their SM values 
and the sign of $\kappa_{\PQt}$ is taken to be positive~\cite{Sirunyan:2018lzm}.
The weak dependence of $\mathcal{D}_{0-}$ distributions for the $\cPqt\PH$ events is neglected. Studies show that it decreases the sensitivity by 0.1 $\,\sigma$. The other processes are constrained to their SM predictions. 

The fit results are shown in Fig.~\ref{fig:bdt_cp} and are obtained using the profile likelihood method
as $\fcp=0.00 \pm 0.33$, with the constraint $\abs{\fcp}<0.67$ at 95\% confidence level (\CL). 
The coverage was determined with pseudo-datasets and found to agree with that expected in the asymptotic limit~\cite{Feldman:1997qc}. 
The pure pseudoscalar model of CP structure of the $\Htt$ coupling ($\fcp=1$) is excluded at 3.2$\,\sigma$.
The expected constraints based on SM simulation are $\fcp=0.00 \pm 0.49$ at 68\% \CL, 
$\abs{\fcp}<0.82$ at 95\% \CL, and 2.6$\,\sigma$ exclusion of the $\fcp=1$ model. 

To conclude, we presented the first single-channel observation of the \ttH process 
and the first measurement of the CP structure of the \Htt coupling using the \Hgammagamma channel. 
The cross section of the \ttH process is measured to be \sigtimesbr$ = 
1.56^{+0.34}_{-0.32}$\unit{fb}, corresponding to $1.38^{+0.36}_{-0.29}$ times the SM prediction, with a significance of $6.6\sigma$.
The data disfavor the pure CP-odd model of the \Htt coupling at $3.2\sigma$, and a possible fractional
CP-odd contribution is constrained to be $\fcp=0.00\pm0.33$ at 68\% \CLnp.

\begin{acknowledgments}
We congratulate our colleagues in the CERN accelerator departments for the excellent performance of the LHC and thank the technical and administrative staffs at CERN and at other CMS institutes for their contributions to the success of the CMS effort. In addition, we gratefully acknowledge the computing centers and personnel of the Worldwide LHC Computing Grid for delivering so effectively the computing infrastructure essential to our analyses. Finally, we acknowledge the enduring support for the construction and operation of the LHC and the CMS detector provided by the following funding agencies: BMBWF and FWF (Austria); FNRS and FWO (Belgium); CNPq, CAPES, FAPERJ, FAPERGS, and FAPESP (Brazil); MES (Bulgaria); CERN; CAS, MoST, and NSFC (China); COLCIENCIAS (Colombia); MSES and CSF (Croatia); RPF (Cyprus); SENESCYT (Ecuador); MoER, ERC IUT, PUT and ERDF (Estonia); Academy of Finland, MEC, and HIP (Finland); CEA and CNRS/IN2P3 (France); BMBF, DFG, and HGF (Germany); GSRT (Greece); NKFIA (Hungary); DAE and DST (India); IPM (Iran); SFI (Ireland); INFN (Italy); MSIP and NRF (Republic of Korea); MES (Latvia); LAS (Lithuania); MOE and UM (Malaysia); BUAP, CINVESTAV, CONACYT, LNS, SEP, and UASLP-FAI (Mexico); MOS (Montenegro); MBIE (New Zealand); PAEC (Pakistan); MSHE and NSC (Poland); FCT (Portugal); JINR (Dubna); MON, RosAtom, RAS, RFBR, and NRC KI (Russia); MESTD (Serbia); SEIDI, CPAN, PCTI, and FEDER (Spain); MOSTR (Sri Lanka); Swiss Funding Agencies (Switzerland); MST (Taipei); ThEPCenter, IPST, STAR, and NSTDA (Thailand); TUBITAK and TAEK (Turkey); NASU (Ukraine); STFC (United Kingdom); DOE and NSF (USA). 
\end{acknowledgments}

\textit{Note added.} -- After we submitted this letter the ATLAS Collaboration submitted the results of a similar study~\cite{PhysRevLett.125.061802}.

\bibliography{auto_generated} 
\cleardoublepage \appendix\section{The CMS Collaboration \label{app:collab}}\begin{sloppypar}\hyphenpenalty=5000\widowpenalty=500\clubpenalty=5000\input{HIG-19-013-authorlist.tex}\end{sloppypar}
\end{document}

%% file: HIG-19-013-authorlist.tex
\vskip\cmsinstskip
\textbf{Yerevan Physics Institute, Yerevan, Armenia}\\*[0pt]
A.M.~Sirunyan$^{\textrm{\dag}}$, A.~Tumasyan
\vskip\cmsinstskip
\textbf{Institut f\"{u}r Hochenergiephysik, Wien, Austria}\\*[0pt]
W.~Adam, F.~Ambrogi, T.~Bergauer, M.~Dragicevic, J.~Er\"{o}, A.~Escalante~Del~Valle, M.~Flechl, R.~Fr\"{u}hwirth\cmsAuthorMark{1}, M.~Jeitler\cmsAuthorMark{1}, N.~Krammer, I.~Kr\"{a}tschmer, D.~Liko, T.~Madlener, I.~Mikulec, N.~Rad, J.~Schieck\cmsAuthorMark{1}, R.~Sch\"{o}fbeck, M.~Spanring, W.~Waltenberger, C.-E.~Wulz\cmsAuthorMark{1}, M.~Zarucki
% ...
\vskip\cmsinstskip
\dag: Deceased\\
1:  Also at Vienna University of Technology, Vienna, Austria\\
% ...
